# Tunable pure spin supercurrents and the demonstration of a superconducting spin-wave device

Kun-Rok Jeon,[1,2]* Xavier Montiel,[3] Sachio Komori,[1] Chiara Ciccarelli,[2] James Haigh,[4] Hidekazu Kurebayashi,[5] Lesley F. Cohen,[6] Chang-Min Lee,[1] Mark G. Blamire[1] and Jason W. A. Robinson[1]*

[1]*Department of Materials Science and Metallurgy, University of Cambridge, 27 Charles Babbage Road, Cambridge CB3 0FS, United Kingdom*

[2]*Cavendish Laboratory, University of Cambridge, Cambridge CB3 0HE, United Kingdom*

[3]*Department of Physics, Royal Holloway, University of London, Egham Hill, Egham, Surrey TW20 0EX, United Kingdom*

[4]*Hitachi Cambridge Laboratory, J. J. Thomson Avenue, Cambridge CB3 0HE, United Kingdom*

[5]*London Centre for Nanotechnology and Department of Electronic and Electrical Engineering at University of College London, London WC1H 01H, United Kingdom*

[6]*The Blackett Laboratory, Imperial College London, SW7 2AZ, United Kingdom*

*To whom correspondence should be addressed: jeonkunrok@gmail.com, jjr33@cam.ac.uk

**Recent ferromagnetic resonance experiments and theory of Pt/Nb/Ni$_8$Fe$_2$ proximity-coupled structures strongly suggest that spin-orbit coupling (SOC) in Pt in conjunction with a magnetic exchange field in Ni$_8$Fe$_2$ are the essential ingredients to generate a pure spin supercurrent channel in Nb. Here, by substituting Pt for a perpendicularly magnetized Pt/Co/Pt spin-sink, we are able to demonstrate the role of SOC, and show that pure spin supercurrent pumping efficiency across Nb is** **tunable by controlling the magnetization direction of Co. By inserting a Cu spacer with weak SOC between Nb and Pt/(Co/Pt) spin-sink, we also prove that Rashba-type SOC is key for forming and transmitting pure spin supercurrents across Nb. Finally, by engineering these properties within a single multilayer structure, we demonstrate a prototype superconducting spin-wave (SW) device in which lateral SW propagation is gateable via the opening or closing of a vertical pure spin supercurrent channel in Nb.**

Spin-triplet Cooper pairs carry a net spin component in addition to charge and are therefore key to the development of superconducting spintronics[1-3], underlying a future revolution in energy-efficient computing. It is established that spin-polarized triplet pairs are generated via spin-mixing and spin-rotation processes at magnetically-inhomogeneous superconductor/ferromagnet (SC/FM) interfaces[1-3]. Recently, theoretical[4-8] and experimental studies[9-13] have been dedicated to an alternative mechanism for triplet pair creation involving spin-orbit coupling (SOC) in combination with a magnetic exchange field $h_{ex}$. In such systems, triplet pair creation depends on the commutation relationship[4-7] between SOC and $h_{ex}$.

The latter mechanism via SOC in conjunction with $h_{ex}$ offers a conceptually novel approach to tune superconducting spin currents, as we demonstrate here using ferromagnetic resonance (FMR) spin pumping[9,14]. When a perpendicularly magnetized Pt/Co/Pt spin-sink is proximity-coupled to Nb (singlet SC) (Fig. 1a), the Co thickness $t_{Co}$-dependent magnetization anisotropy[15,16] changes its effective tilt angle $\Theta_{Co}$ under in-plane (IP) FMR condition of the IP magnetized $Ni_8Fe_2$ (Fig. 1b). This in turn alters the degree of orthogonality between $h_{ex}$ and SOC at the interface of Nb and Pt/(Co/Pt) spin-sink.

Manipulating $\Theta_{Co}$ determines the efficiency with which spin-zero triplets (converted from spin singlets by the presence of $h_{ex}$) rotate to form equal-spin triplets[4-6]. This enables orthogonality tuning of spin-angular-momentum transfer from the precessing $Ni_8Fe_2$ through the proximity-induced equal-spin triplets into singlet Nb layers, which we call superconducting pure spin currents[9]. Such transmitted spin currents to Pt/Co/Pt spin-sinks result in the enhanced spin pumping/transfer which is then probed by measuring the FMR linewidth broadening or Gilbert damping increase of the middle $Ni_8Fe_2$ layer[9,14].

To demonstrate our approach, we perform a series of FMR measurements on Pt/Co/Pt/Nb/$Ni_8Fe_2$/Nb/Pt/Co/Pt multilayers (Fig. 1a). The ultrathin ($\leq 1.5$ nm) perpendicularly magnetized Co layers serve as an internal source of $h_{ex}$ to the neighbouring (inner) Pt layers, supplying spontaneous spin splitting[17,18] with out-of-plane (OOP) polarization (Fig. 1a); the outer Pt layers boost the perpendicular anisotropy of the Co as well as the total effective spin conductance of Pt/Co/Pt trilayers[19]. By inserting a thin Cu spacer with weak SOC at the interface between Nb and Pt/(Co/Pt) layers (Fig. 1a), we are able to separate the contribution of interfacial Rashba-type SOC to the $\theta_{Co}$-dependent superconducting spin-pumping efficiency from other contributions and to compare it with the prediction from spin-triplet proximity theory[4-6,8].

We first measure the $t_{Co}$-dependence of the superconducting transition $T_c$ (Fig. 1c) for a series of multilayers with and without Cu spacers. $T_c$ decreases rapidly with increasing $t_{Co}$ until it reaches about 1.5 nm where it slightly increases. No significant change in $T_c(t_{Co})$ appears with the addition of the Cu spacer, consistent with its long (thermal) coherence length of several hundred nanometers[3]. In analogy with the original consideration on the non-uniform superconducting state[20,21], such non-monotonic $T_c$ behaviour has been discussed based on a spatial modulation of the superconducting order parameter due to Cooper pairs acquiring a non-zero net momentum in the presence of $h_{ex}$; in particular, for SC/FM multilayers or SC/FM bilayers with FM thickness $t_{FM}$ of the order of the coherence length $\xi_{FM}$, which leads to a damped oscillatory behaviour of the order parameter[22]. A quantitative analysis (see Methods) of the $T_c$ data (black lines, Fig. 1c) gives an effective $\xi_{FM}$ of 1.4−1.6 nm and interface transparency $\gamma_B = 0.18-0.20$ for our samples, which are in reasonable agreement with those obtained from Nb/FM[23] bilayers and Nb/Cu/FM trilayers[24] with strong FMs.

The $t_{Co}$-dependent magnetization anisotropy of the Pt/Co/Pt spin sinks can be independently characterized by static magnetometry measurements on Pt/Co/Pt/Nb-only films with different $t_{Co}$. Figure 1e shows the typical magnetization hysteresis $m(H)$ curves obtained at 8 K by applying the external magnetic field $\mu_0H$ parallel and perpendicular to the film plane. At low $t_{Co}$ ($\leq 0.8$ nm), the easy axis of the Co magnetization $M_{Co}$ is clearly OOP, indicating that the ultrathin Co sandwiched between two Pt layers has well-established perpendicular magnetization anisotropy (PMA), as expected for the Pt 5d−Co 3d orbital hybridization at either Pt/Co interface plus SOC[15]. As $t_{Co}$ approaches 1.5 nm, the predominant magnetization anisotropy changes from OOP to IP, exhibiting the reorientation transition[16]. Using the relationship[16] $\mu_0H_{ani} \cdot M_s/2 = K_{eff}$, where $\mu_0H_{ani}$ is the anisotropy field and $M_s$ is the saturation magnetization, the effective PMA energy $K_{eff}$ is estimated for $t_{Co} \leq 0.8$ nm to be ~ 1 MJ m$^{-3}$, comparable to typical values of the perpendicularly magnetized Pt/Co/Pt trilayers[25].

Assuming coherent rotation of $M_{Co}$ from OOP under the application of IP resonance fields $\mu_0H_{res}$ for the middle Ni$_8$Fe$_2$, the effective $\theta_{Co}$ can be estimate using the simple Stoner-Wohlfarth model where $\theta_{Co} = \arccos[M(\mu_0H_{res})/M_s]$. We then achieve discrete tilt states of the Pt/Co/Pt spin sinks from OOP to IP (Fig. 1f), which are systematically controllable by varying $t_{Co}$.

We next show the influence of the tilt states on the superconducting spin-pumping efficiency, namely that the associated orthogonality between $h_{ex}$ and SOC at the Nb/Pt/(Co/Pt) interface strongly modifies the spin-angular-momentum transfer in the superconducting state. Figure 2a,b shows the microwave frequency $f$ dependence of FMR data for the Cu-absent (Cu-present) samples, taken above and below $T_c$ of the Nb layers. From this, we extract the effective Gilbert damping $\alpha$, which provides a measure[9,14,18] of the net spin current flow out of the precessing $Ni_8Fe_2$, and the effective saturation magnetization $\mu_0 M_s$ (see Methods).

The extracted $\alpha$ and $\mu_0 M_s$ values are plotted as a function of $t_{Co}$ in Fig. 2c. In the normal state ($T/T_c > 1$), $\alpha$ is almost $t_{Co}$-independent for both sample sets and there is a small decrease in the magnitude by introducing Cu spacers. This means that the presence of ultrathin Co($\leq$ 2 nm) and Cu(5 nm) layers hardly changes the normal-state spin-pumping behaviour, as expected from their small spin conductances[19] relative to Pt and the three layers (Co, Cu, Pt) are all approximately spin-transparent[26] with each other due to their similar crystal and electronic structures.

However, in the superconducting state ($T/T_c < 1$), a significant $t_{Co}$-dependent enhancement of $\alpha$ appears and is strongly affected by the addition of Cu. For the Cu-absent multilayers, as $t_{Co}$ increases the superconducting-state damping enhancement (indicating the enhanced spin flow/transfer mediated most likely by equal-spin triplet pairing)[8,14,18] rapidly rises until reaching 0.8 nm and then slowly decreases for thicker Co layers, resulting in a maximum at $t_{Co} \approx 0.8$ nm. For the Cu-present samples, the overall amplitude of damping enhancement diminishes compared with the Cu-absent samples and the maximum moves to a lower value of $t_{Co}$ (0.4 nm). Since this non-trivial enhancement of $\alpha(t_{Co})$ occurs in the ultrathin regime ($t_{Co} \leq 2$ nm, about one order of magnitude smaller than the spin diffusion length[27]) only for the superconducting state, it must reflect how the tilt states of the Pt/Co/Pt spin sinks correlate with the superconducting spin transport.

To elucidate this, we plot the damping difference across $T_c$, defined as $[\alpha_{2\,K} - \alpha_{8\,K}]/2\Delta_{2\,K}$ where $2\Delta$ is the superconducting gap at 2 K calculated from the measured $T_c$ (Fig. 1d), with and without Cu spacers as a function of the effective $\theta_{Co}$ (Fig. 2e). In the absence of the Cu spacers, $[\alpha_{2\,K} - \alpha_{8\,K}]/2\Delta_{2\,K}$ rapidly rises with increasing $\theta_{Co}$ from 0° to 56° followed a fall for a higher angle. However, this characteristic angular dependence vanishes when the Cu spacer (with weak SOC) is present: the damping difference increases monotonically/slowly up to the highest angle and it saturates to the similar value of the Cu-absent $\theta_{Co} \approx 76°$ ($t_{Co} = 0.4$ nm) sample.

There are, in principle, two different sources of proximity-induced triplet pairing which can contribute to the characteristic angular dependence observed in our experimental setup. Firstly, it is well-known that magnetization noncollinearity (or inhomogeneity)[1-3,28] between two FMs separated by a SC of the order of the coherence length can generate equal-spin triplets through the entire structure. The equal-spin triplet density is then ascribed to the relative magnetization angle $\theta$ between the two FMs[28]: $\propto \boldsymbol{M}_1 \times \boldsymbol{M}_2 \propto \sin(\theta)$. This explains why our $\theta_{Co} \approx 76°$ ($t_{Co} = 0.4$ nm) samples reveal larger enhancements than $\theta_{Co} \approx 9°$ ($t_{Co} = 2.0$ nm) samples (Fig. 2e). Secondly, even for a single magnetically-homogeneous FM, the equal-spin triplet correlation can be generated by introducing a strongly SO coupled interface (e.g. Pt) between the FM and SC[4-6,8]. In this case, the singlet-triplet conversion efficiency is predicted to scale with the degree of orthogonality between SOC and $h_{ex}$; or equivalently, the cross product of the SO vector operator $\left[\hat{A}_k, \left[\hat{A}_k, h^a\sigma^a\right]\right]$ and the exchange field operator $h^a\sigma^a$. Here $\hat{A}_{k=x,y,z}$ is the vector potential describing the form of the SOC, for instance, the Rashba constant $\alpha_R$ (Dresselhaus constant $\beta_D$) due to the interface (bulk) inversion asymmetry. $\sigma^a (h^a)$ with $a =$ x,y,z is the vector of Pauli matrices (exchange field).

For a metallic vertical structure with *atomically flat* interfaces, the vector potential can be approximated as[4,5]: $\hat{A}_x \approx 0$, $\hat{A}_y \approx -\beta_D \sigma^y + \alpha_R \sigma^z$, $\hat{A}_z \approx \beta_D \sigma^z - \alpha_R \sigma^y$. With finite Rashba ($\alpha_R \neq 0$) and zero Dresselhaus ($\beta_D = 0$) contribution to the SOC[6,8], as relevant to our experimental setup, a sinusoidal maximum of the equal-spin triplet correlation is expected when the canting angle between IP and OOP components of $h_{ex}$ becomes $45^0$. In such a case, the overall triplet density is quadratic in $\alpha_R$ and so very sensitive to details of the spin-orbit coupled interface. The addition of a-few-nm-thick Cu spacer layers[26,29] at the spin-orbit coupled interface turns out to be sufficient to significantly quench the interfacial Rashba-type SOC, thus providing an important test experiment for the responsible mechanism, being done clearly here (inset of Fig. 2e,f).

We also note that non-vanishing of $\hat{A}_x$ ($\neq 0$), as would be expected from *non-ideal* interfaces where OOP component of the Rashba SO field with respect to the local interface plane survives on a scale of the coherence length[7], allows the equal-spin triplet to be generated locally even with a purely IP magnetized FM ($h^x = 0$). Each triplet channel is then able to transport spin-angular-momentum from the precessing FM ($Ni_8Fe_2$) through a singlet SC (Nb) to a spin dissipative bath (Pt spin-sink) independently even if the spatial average of net polarization of total triplet channels over the entire interface plane becomes zero. This is a likely mechanism for our previous FMR experiments[9,18] and for the $t_{Co} = 0$ samples (Fig. 2e). When the Pt spin-sink is substituted for the perpendicularly magnetized Pt/Co/Pt spin-sink, a global triplet channel opens in addition to the local channels, maximizing the overall superconducting spin-pumping efficiency at $\theta_{Co} \approx 45°$ (Fig. 2e).

By taking all these effects together, we can arrive at good fits to $[\alpha_{2\,K} - \alpha_{8\,K}]/2\Delta_{2\,K}$ vs. $\theta_{Co}$ data for both sample sets (black solid and dashed lines in Fig. 2e, see Supplementary Section 1,2), thereby reproducing the experimental results and capturing the underlying physics. To focus on the second SOC mechanism, in particular for the interfacial contribution, we take the difference between the data with and without the Cu spacer (Fig. 2f). We then find an approximately sinusoidal maximum at $\theta_{Co} \approx 45°$, which is in good agreement with the Rashba SOC-induced triplet pairing[6,8] described above. The data described above provides a proof-of-concept result demonstrating the orthogonality tuning of superconducting spin currents.

To help understand better the FMR absorption data of symmetric structures (Fig. 2), we also measure the $t_{Co}$-dependence of spin-pumping-induced inverse spin-Hall effect (iSHE)[30,31] for the additional sets of asymmetric Pt/Co/Pt/Nb/Ni$_8$Fe$_2$ structures with and without Cu spacers. This provides direct evidence for spin transport in the normal state. Figure 3a (3b) displays the transverse d.c. voltage signals vs. external IP $\mu_0 H$ for the Cu-absent (Cu-present) samples at $f = 5$ GHz, taken above and below $T_c$ (see Methods). Under IP FMR of the Ni$_8$Fe$_2$, a clear Lorentzian peak emerges in the dc voltage only in the normal state for both sample sets, which can be explained[31] by the strong decay of the quasiparticle charge-imbalance relaxation length across $T_c$. Importantly, the polarity of the Lorentzian peak is identical (opposite) to that of Pt/Ni$_8$Fe$_2$ (Nb/Ni$_8$Fe$_2$) bilayers[31], where the Pt (Nb) spin sink is known to have a positive (negative) spin-Hall angle $\theta_{SH}$[19,31]. This indicates that the pumped spin currents from the precessing Ni$_8$Fe$_2$ pass through the Nb(30 nm) layer to a large extent to the (Cu)/Pt/Co/Pt spin sinks and the overall iSHE in our structures is dominated by the (Cu)/Pt/Co/Pt (rather than the Nb).

For a quantitative analysis, we plot the iSHE voltage divided by sample resistance $V_{iSHE}/R$ vs. $t_{Co}$ (Fig. 3c) and $\theta_{Co}$ (Fig. 3d). In these plots, we can see that there is a clear decrease in the iSHE signal by the addition of Cu and its magnitude is strongly $\theta_{Co}$-dependent, which can be described by the rapid spin precession/dephasing of transverse spins[32] around $h_{ex}$ of the Co layer: $\cos^2(\theta_{Co})$ (black lines in Fig. 3d). Note that the signal difference caused by the addition of 5 nm of Cu (insets in Fig. 3c,d) is nearly $\theta_{Co}$-independent. These results taken together support our argument that Cu spacers weaken the interfacial SOC strength and it is the Co tilt state that then plays a dominant role in the spin transport process.

Finally, we progress to show the potential to harness these effects in a proof-of-principle prototype superconducting SW device (Fig. 4). The idea behind this is that lateral SW propagation[33,34] in our proximity-engineered structure (e.g. $\alpha_{2\,K} - \alpha_{8\,K} \approx 0.005$ for the $t_{Co} = 0.8$ nm sample) between microwave injector and detector antennas is readily altered by opening or closing the vertical spin transport channel via the proximity creation of triplet pairing. Figure 4a,b shows the $f$-dependent SW transmission $\Delta S_{12}$ of two types of the SW devices with and without Pt/Co(0.8 nm)/Pt spin sinks, obtained above and below $T_c$ at the fixed/external IP $\mu_0 H = 70$ mT in the magnetostatic surface wave (MSSW) geometry[33,34] (see Methods and Supplementary Section 6 for details). The observed spectra containing two major peaks in the low $f$ (< 7 GHz) regime and satisfying the SW dispersion relationship (Supplementary Video 1-4 and Supplementary Section 6) and their exponential decay in the intensity with increasing the distance $d$ between the two separate antennas (Fig. 4c,d) indicate the propagating SWs[33,34].

The most noteworthy aspect in this demonstration is that on entering the superconducting state, the intensity of lateral SW transmission signal clearly rises (decays) when the Pt/Co(0.8 nm)/Pt spin sinks are omitted (added) [Fig. 4c (4d), see also Supplementary Video 1-4] and the degree of this change becomes pronounced with increasing $d$. This is because SWs experience weaker (stronger) effective attenuation during laterally propagating if spin-angular-momentum is less (more) transmitted across the adjacent/superconducting Nb to the spin loss regimes in vertical direction. Note that the SW attenuation increases proportionally to the total FMR damping of the system[33,34].

With the Pt/Co(0.8 nm)/Pt spin sinks, we are able to modulate the lateral SW transmission intensity up to about 40% by proximity-generating the vertical triplet spin-transport channel. This result is encouraging and may provide a new type of SW logic functionality[35] activated in the superconducting state.

We have shown that when a perpendicularly magnetized Pt/Co/Pt spin sink is proximity-coupled to Nb, superconducting spin-pumping efficiency can be tuned by controlling the effective $\theta_{Co}$ - i.e. by tuning the degree of orthogonality between the SOC and $h_{ex}$ at the Nb/Pt/(Co/Pt) interface[4-6,8]. We have also found that by comparison with the Cu-present samples, the $\theta_{Co}$-dependent superconducting spin-pumping efficiency reflects characteristic features of Rashba SOC-induced triplet pairing[4-6,8]. Our results provide a timely step towards understanding key interfacial properties for tuning superconducting spin transport mediated via equal-spin triplet states in a spin-singlet superconductor. Our finely proximity-engineered structures enable experimental realization of a prototype superconducting SW device. This concept can be extended to any Rashba system[36,37] for the development of superconducting spin-logic devices[1] in which SOC is gate-tunable[36], leading to a superconducting spin-based transistor.

## Methods

Methods and any associated references are available in the online version of the paper.

## Acknowledgements

This work was supported by EPSRC Programme Grant EP/N017242/1.

## Author contributions

K.-R.J. and J.W.A.R. conceived and designed the experiments. The samples were prepared by K.-R.J. and the magnetization and transport measurements were performed by K.-R.J. with help from S.K. The SW devices were fabricated by K.-R.J., C.-M.L. and S.K. The FMR absorption and iSHE measurements were carried out by K.-R.J. with help of C.C. The SW transmission spectra were measured by K.-R.J. with the guidance of J.H. The data analysis was performed by K.-R.J., C.C., J.H., H.K., L.F.C., M.G.B., S.K. and J.W.A.R. The model calculation was undertaken by X.M. All authors discussed the results and commented on the manuscript, which was written by K.-R., X.M. and J.W.A.R.

## Additional information

Supplementary information is available for this paper here (TBD). Reprints and permissions information is available at www.nature.com/reprints. Correspondence and requests for materials should be addressed to K.-R.J. or J.W.A.R.

## Competing interests

The authors declare no competing financial interests.

## Methods

**Sample growth.** Symmetric Pt/Co/Pt/Nb/Ni$_8$Fe$_2$/Nb/Pt/Co/Pt and asymmetric Pt/Co/Pt/Nb/Ni$_8$Fe$_2$ multilayers, with and without Cu spacer layers, were grown on 5 mm × 5 mm thermally oxidized Si substrates by d.c. magnetron sputtering in an ultra-high vacuum chamber[9,18]. The symmetric and asymmetric structures were prepared, respectively, for the ferromagnetic resonance (FMR) absorption[9,18] and inverse spin-Hall effect (iSHE, or transverse d.c. voltage)[31] measurements. All layers were grown *in-situ* at room temperature. $Ni_8Fe_2$, Nb, Co and Cu are deposited at an Ar pressure of 1.5 Pa and Pt at 3.0 Pa. The typical deposition rates were 5.1 nm/min for $Ni_8Fe_2$, 21.1 nm/min for Nb, 6.0 nm/min for Co, 9.7 nm/min for Cu and 7.6 nm/min for Pt. The thicknesses of $Ni_8Fe_2$, Nb, inner (outer) Pt and Cu layers were kept constant at 6, 30, 1.7 (2.2) and 5 nm, respectively, while the thickness of the Co layer varied from 0 to 2 nm to investigate the variation of FMR damping as a function of $t_{Co}$ (or the Co tilt angle $\theta_{Co}$) through the superconducting transition temperature $T_c$ of the coupled Nb. Note that for all samples, the Nb (inner Pt) thickness is fixed at 30 (1.7) nm where the Pt/Co/Pt spin sink was proximity-coupled through the Nb layer to the precessing $Ni_8Fe_2$ layer and the largest enhancement of spin pumping in the superconducting state was achieved in our prior FMR experiments[9,18].

**Magnetization characterization.** The static magnetization hysteresis curves were measured on 5 mm × 5 mm samples using a Quantum Design Magnetic Property Measurement System at 8 K, immediately above the superconducting transition temperature $T_c$. The external magnetic field was applied parallel and perpendicular to the film plane direction.

**Superconducting transition measurement.** d.c. electrical transport measurements were conducted on (un-patterned) 5 mm × 5 mm samples using a custom-built dipstick probe in a liquid helium dewar with a four-point current-voltage method. The resistance $R$ (of a sample) vs. temperature $T$ curves were obtained at the applied current $I$ of ≤ 0.1 mA while decreasing $T$. From the $T$ derivative of $R$, $dR/dT$, $T_c$ was defined as the $T$ value that exhibits the maximum of d$R$/d$T$.

We analyzed our $T_c(t_{Co})$ data (Fig. 1d) using the following approximate formula[22]:

$\ln\left[\frac{T_c^*}{T_c}\right] \approx \Psi\left(\frac{1}{2}\right) - Re\left\{\Psi\left[\frac{1}{2} + \frac{2T_c}{T_c^*\tilde{\tau}_0} \times \frac{1}{\tilde{\gamma} + \frac{1-i}{2}\cosh[(1+i)t_{FM}]}\right]\right\}$, where $T_c^* = T_c(t_{FM} = 0)$, $\Psi$ is the digamma function, $\tilde{\tau}_0^{-1} = (1/4\pi T_c)(D_{SC}/t_{SC}\xi_{FM})(\rho_{SC}/\rho_{FM})$, $D_{SC}$ is the diffusion coefficient of the Nb (10 cm$^2$/s at 8 K), $t_{SC}$ is the Nb thickness (30 nm) and $\rho_{SC}$ ($\rho_{FM}$) is the conductivity of the Nb (Co) [7 (30) $\mu\Omega$-cm at 8 K]. $\tilde{\gamma} = \gamma_B(\xi_{SC}/\xi_{FM})$, $\gamma_B$ is the interface transparency and $\xi_{SC}$ is the (dirty-limit) coherence length of the Nb (16–18 nm at 2 K)[9]. Note that in this formula, only the influence of $h_{ex}$ on the order parameter is taken into account[22].

**Broadband FMR absorption and iSHE measurements.** We measured the FMR response of the sample attached on a broadband coplanar waveguide (CPW) with either d.c. field or r.f. pulse modulation[9,18]. To obtain each FMR spectrum, the microwave power absorbed by the sample was measured while sweeping the external static magnetic field $\mu_0 H$ at the fixed microwave frequency $f$ of 5–20 GHz. At the beginning of each measurement, we applied a large IP $\mu_0 H$ (0.5 T) to fully magnetize the Ni$_8$Fe$_2$ layer, after which the field was reduced to the range of FMR. Once the $f$-dependent FMR measurements (from high to low $f$) were complete, the field was returned to zero to cool the system down further for a lower $T$ measurement. For all FMR absorption measurements, the microwave (MW) power was set to 10 dBm where the actual microwave power absorbed in the sample is a few mW that has no effect on $T_c$ of the Nb layer[9]. Note also that the fixed thickness (30 nm) of Nb layers studied here is much less than the magnetic penetration depth in the superconducting state ($\geq$ 100 nm in thin Nb films) and so there is no considerable effect of Meissner screening on the local (d.c./r.f.)

magnetic field experienced by $Ni_8Fe_2$ below $T_c$, as supported by the insensitivity of the resonance field $\mu_0 H_{res}$ across $T_c$ (Fig. 2a,b). We employed a vector field cryostat from Cryogenic Ltd. that can apply a 1.2-T-magnetic field in any direction over a $T$ range of 2–300 K.

We first fitted the FMR absorption data (Supplementary Section 3) with the field derivative of symmetric and antisymmetric Lorentzian functions[38] to accurately determine the FMR linewidth $\mu_0 \Delta H$ and the resonance field $\mu_0 H_{res}$: $\frac{d\chi"}{dH} \propto A \cdot \left[\frac{(\Delta H_{HWHM})^2 \cdot (H-H_{res})}{[(\Delta H_{HWHM})^2 + (H-H_{res})^2]^2}\right] + B \cdot \left[\frac{(\Delta H_{HWHM}) \cdot (H-H_{res})^2 - (\Delta H_{HWHM})^3}{[(\Delta H_{HWHM})^2 + (H-H_{res})^2]^2}\right]$, where $A$ ($B$) is the amplitude of the field derivative of the symmetric (antisymmetric) Lorentzian function, $\mu_0 H$ is the external d.c. magnetic field and $\mu_0 \Delta H_{HWHM} = \frac{\sqrt{3}}{2} \mu_0 \Delta H$ is the half-width-at-half-maximum (HWHM) of the imaginary part $\chi"$ of the magnetic susceptibility.

From the linear scaling of $\mu_0 \Delta H$ with $f$ (Fig. 2a,b), we calculated the effective Gilbert(-type) damping constant $\alpha$: $\mu_0 \Delta H(f) = \mu_0 \Delta H_0 + \frac{4\pi \alpha f}{\sqrt{3}\gamma}$, here $\mu_0 \Delta H_0$ is the zero-frequency line broadening. We also estimated the effective saturation magnetization $\mu_0 M_s$ (of the $Ni_8Fe_2$) from the dispersion relation of $\mu_0 H_{res}$ with $f$ (inset of Fig. 2a,b) using Kittel's formula: $f = \frac{\gamma}{2\pi}\sqrt{[\mu_0(H_{res} + M_{eff}) \cdot \mu_0 H_{res}]}$, where $\gamma = g_L \mu_B/\hbar$ is the gyromagnetic ratio ($1.84 \times 10^{11}$ $T^{-1}$ $s^{-1}$), $g_L$ is the Landé g-factor (taken to be 2.1), $\mu_B$ is the Bohr magneton and $\hbar$ is Plank's constant divided by $2\pi$.

For the iSHE (or transverse d.c. voltage) measurement[31], the sample was attached face down on the CPW by using an electrically insulating high-vacuum grease. A microwave signal was passed through the CPW and excited FMR of the $Ni_8Fe_2$ layer; a transverse d.c. voltage as a function of $\mu_0 H$ was measured between two Ag-paste contacts at opposite ends of the sample. In these measurements, the microwave frequency was fixed at 5 GHz and the microwave power at the CPW at approximately 150 mW (for $T =$ 2 and 8 K), which yields measurable signals (≥ 5 nV) in our setup.

The measured d.c. voltage (Fig. 3a,b) can be decomposed into symmetric and antisymmetric Lorentzian functions with respect to $\mu_0 H_{res}$, with weights of $V_{sym}$ and $V_{asy}$ respectively[31]: $V(H) - V_0 = V_{sym} \cdot \left[\frac{(\Delta H')^2}{(\Delta H')^2 + (H-H_{res})^2}\right] + V_{asy} \cdot \left[\frac{(\Delta H') \cdot (H-H_{res})}{(\Delta H')^2 + (H-H_{res})^2}\right]$, where $V_0$ is a background voltage and $\mu_0 \Delta H'$ is the HWHM of the d.c. voltage. We attributed $V_{sym}$ to the iSHE signal $V_{iSHE}$. If the Co thickness in the Pt/Co/Pt spin sink is larger than its spin dephasing length (a few ångstroms)[32], $V_{iSHE}(\theta_{Co})$ is simply proportional to $\cos^2(\theta_{Co})$ (Fig. 3d).

**SW device fabrication.** To fabricate the standard SW devices[33,34] displayed in Supplementary Section 6, the Hall bar(-type) structures with an active SW track of 50 × 50 μm² were patterned into the *in-situ* grown Nb/Ni$_8$Fe$_2$/Nb films with and without Pt/Co(0.8 nm)/Pt spin sinks by using optical lithography and Ar-ion beam etching. After depositing AlN(40 nm) for d.c. electrical isolation by reactive sputtering, coplanar waveguides (CPWs or, MW antennas) with a various inter-spacing of 10–25 μm were patterned on top of the SW track using electron-beam lithography and lift-off of sputtered Cu(100 nm)/Ti(5 nm) layers. Two identical CPWs consist of a MW signal line (2 μm wide) and two ground lines (1 μm wide) with an intra-separation of 2 μm, which preferentially excites or detects the SWs with a wavenumber $k_{SW}$ in the range of 0.9 ± 0.6 μm⁻¹ (Supplementary Section 6).

**Propagating SW spectroscopy.** A pair of antennas of the SW device were connected to ports 1 and 2 of a vector network analyser (VNA, Rohde & Schwarz, 100 MHz to 20 GHz) by multiple wire bonding to a pre-calibrated sample holder (having the 50 Ω impedance) via phase-stable coaxial cables. The $f$-dependent forward complex transmission coefficient (e.g. scattering parameter $S_{12}$: the MW power received at port 1 relative to the power conveyed to port 2) was measured in the variable temperature insert of a vector field cryostat by applying a fixed/external IP $\mu_0 H$ transverse to the SW propagation direction (or wave vector $\boldsymbol{k}_{SW}$), so-called the MSSW configuration[33,34]. The input MW power was set to 0 dBm (the actual power delivered to the device in our setup: < 100 µW) so that the non-linear response of magnetization dynamics and the unintentional heating effect on the Nb layer can be avoided. The SW transmission signal $\Delta S_{12}(f, \mu_0 H)$ of interest was analyzed by subtracting the non-magnetic background $S_{12}(f, \mu_0 H_{ref})$ under application of a large reference field $\mu_0 H_{ref}$ (0.12 T) and normalizing the $f$-dependence[33,34]: $\Delta S_{12}(f, \mu_0 H) = \frac{S_{12}(f, \mu_0 H) - S_{12}(f, \mu_0 H_{ref})}{S_{12}(f, \mu_0 H_{ref})}$.

The SW dispersion in the MSSW mode for symmetric sample structures is given by[33,34]: $f_{SW} \approx \frac{\gamma}{2\pi} \sqrt{\left[\mu_0(H_{res} + M_{eff}) \cdot \mu_0 H_{res} + \left(\frac{\mu_0 M_{eff}}{2}\right)^2 (1 - \exp(-2k_{SW}t))\right]}$, where $t$ is the Ni$_8$Fe$_2$ thickness (6 nm). By fitting the SW resonance, corresponding to the peak in the absolute of $\Delta S_{12}$ (= $|\Delta S_{12}|$, Fig. 4a,b), to this dispersion relationship, we extracted the $k_{SW}$ and $\mu_0 M_{eff}$ values (Supplementary Section 6) for the Ni$_8$Fe$_2$ layer. In addition, we deduced the SW attenuation length $\lambda_{SW}$ (insets of Fig. 4c,d) from the fact[33,34] that the SW intensity, defined as the maximum peak-to-valley height of the real part of $\Delta S_{12}$ (= $Re[\Delta S_{12}]$, Fig. 4a,b), exponentially decays with increasing $d$: $\exp\left(-\frac{d}{\lambda_{SW}}\right)$. Here $\lambda_{SW} = \frac{v_g}{\tau_{pres}}$, $v_g = 2\pi \cdot \left(\frac{\partial f_{SW}}{\partial k_{SW}}\right)$ is the group velocity and $\tau_{pres} = \{\alpha\gamma[\mu_0(2H_{res} + M_{eff})]\}^{-1}$ is the magnetization precession time.

## Data availability

The data used in this paper can be accessed here (TBD).

## Code availability

The code that was used for the numerical calculations is available from X.M. upon reasonable request.

## Figure legends

**Figure 1. Principle of the approach and experimental setup a**, Schematic of Pt(2.0 nm)/Co($t_{Co}$)/Pt(1.7 nm)/Nb(30 nm)/Ni$_8$Fe$_2$(6 nm)/Nb(30 nm)/Pt(1.7 nm)/Co($t_{Co}$)/Pt(2.0 nm) multilayers with different Co thicknesses $t_{Co}$. The Cu spacer with weak spin-orbit coupling (SOC) is selected to quench the interfacial Rashba-type SOC at the Nb/Pt/(Co/Pt) interface. **b**, Measurement scheme and Cartensian coordinate system used in the present study. **c**, Normalized resistance $R/R_N$ vs. temperature $T$ plots for three different sets of the samples, grown each in a single deposition run. **d**, $t_{Co}$ dependence of the superconducting transition temperature $T_c$ of the sample sets with and without Cu(5 nm) spacer layers; for comparison, $T_c$ of a bare Nb(30 nm) film is also shown. The black solid (dashed) line is a fit to estimate the effective values of coherence length and interface

transparency (see Methods) for the Cu-absent (Cu-present) samples. **e**, In-plane and out-of-plane magnetization hysteresis $m(H)$ curves of Pt(2.0 nm)/Co($t_{Co}$)/Pt(1.7 nm)/Nb(30 nm)-only films, measured at 8 K. **f**, Effective tilt angle $\theta_{Co}$ of the Co layer estimated from **e** using the Stoner-Wohlfarth model, in which only the corresponding regime of the in-plane $m(H)$ curves to the ferromagnetic resonance measurement condition/sequence (*i.e.* from high to low field, see Methods) is considered.

**Figure 2. Correlation of Co tilt angle with superconducting spin-pumping efficiency.**
**a**, Microwave frequency $f$ dependence of ferromagnetic resonance (FMR) absorption for symmetric Pt(2.0 nm)/Co($t_{Co}$)/Pt(1.7 nm)/Nb(30 nm)/Ni$_8$Fe$_2$(6 nm)/Nb(30 nm)/Pt(1.7 nm)/Co($t_{Co}$)/Pt(2.0 nm) samples with various Co thicknesses, taken above and below $T_c$ of the couple Nb. From this, one can extract the (effective) Gilbert(-type) damping $\alpha$ and the (effective) saturation magnetization $\mu_0 M_s$. **b,** Data equivalent to **a** but for symmetric Pt(2.0 nm)/Co($t_{Co}$)/Pt(1.7 nm)/Cu(5 nm)/Nb(30 nm)/Ni$_8$Fe$_2$(6 nm)/Nb(30 nm)/Cu(5 nm)/Pt(1.7 nm)/Co($t_{Co}$)/Pt(2.0 nm) samples. Note that in any case, the zero-frequency line broadening $\mu_0 \Delta H_0$ due to long-range magnetic inhomogeneities is less than |0.5 mT| and the FMR linewidth $\mu_0 \Delta H$ scales linearly with $f$, indicating the high quality of the samples and the absence of two-magnon scattering[38]. Extracted $\alpha$ (**c**) and $\mu_0 M_s$ (**d**) values as a function of $t_{Co}$ for the samples with and without the Cu spacer. The dashed lines are guide to the eyes. **e**, Damping difference across $T_c$, denoted as $[\alpha_{2\,K} - \alpha_{8\,K}]/2\Delta_{2\,K}$ where $2\Delta$ is the superconducting gap at 2 K calculated from the measured $T_c$ (Fig. 1d), as a function of the (effective) Co tilt angle $\theta_{Co}$. The black solid (dashed) line is a fit from spin-triplet proximity theory[4-6,8] for the Cu-absent (Cu-present) samples (Supplementary Section 1,2). **f**, Interfacial SOC contribution $\Delta[\alpha]_{SOC}$, separated by taking the difference between the $[\alpha_{2\,K} - \alpha_{8\,K}]/2\Delta_{2\,K}$ data (**e**) with and without the Cu spacer. The black solid is a theoretical fit based on Rashba-type SOC-induced triplet paring[6,8] (Supplementary Section 1,2). Here, the amplitude and component of Rashba SO field and the exchange field strength are only adjustable parameters to get to the theoretical fit. The inset of **e,f** shows $[\alpha_{2\,K} - \alpha_{8\,K}]/2\Delta_{2\,K}$ data as a function of Cu spacer thickness $t_{Cu}$ for the $t_{Co}= 0.8$ nm samples (Supplementary Section 4). The red and blue symbols in **c,d** represent independent sets of the samples grown each in a single deposition run.

**Figure 3. Effect of Cu spacer addition and Co tilt angle on normal spin-transport properties.** **a**, Transverse d.c. voltage measurements for asymmetric Pt(2.0 nm)/Co($t_{Co}$)/Pt(1.7 nm)/Nb(30 nm)/Ni$_8$Fe$_2$(6 nm) samples with various Co thicknesses $t_{Co}$ at a fixed microwave frequency $f = 5$ GHz, taken above and below $T_c$ of the couple Nb. The black solid lines are fits to Lorentzian functions (see Methods). **b**, Data equivalent to **a** but for asymmetric Pt(2.0 nm)/Co($t_{Co}$)/Pt(1.7 nm)/Cu(5 nm)/Nb(30 nm)/Ni$_8$Fe$_2$(6 nm) samples. Inverse spin-Hall effect (iSHE) voltage divided by the sample's resistance $V_{iSHE}/R$ as a function of $t_{Co}$ (**c**) and the (effective) Co tilt angle $\theta_{Co}$ (**d**) at $f = 5$ GHz. The dashed lines are guide to the eyes. The left (right) inset in **c** (**d**) shows the signal difference caused by the Cu(5 nm) spacer addition whereas the right inset in **c** exhibits the Cu spacer thickness $t_{Cu}$ dependence of iSHE for the $t_{Co}= 0.8$ nm samples (see Supplementary Section 5 for details). The dashed lines in **c** are guide to the eyes whereas the black solid (dashed) line in **d** is a fit to $\cos^2(\theta_{Co})$ for the Cu-absent (Cu-present) samples. The red and blue symbols in **c,d** represent independent sets of the samples grown each in a single deposition run.

**Figure 4. Experimental realization of superconducting spin-wave devices. a**, Spin-wave (SW) transmission $\Delta S_{12}$ as a function of frequency $f$ for the Nb(30 nm)/Ni$_8$Fe$_2$(6 nm)/Nb(30 nm) device with a different distance $d$ (10−25 µm) between two separate antennas. These spectra are obtained under application of a fixed/external magnetic field $\mu_0 H = 70$ mT above and below $T_c$ of the coupled Nb. In each figure, the red, blue and black curves represent respectively the real, imaginary and absolute of $\Delta S_{12}$. **b**, Data equivalent to **a** but for the Pt(2.0 nm)/Co(0.8 nm)/Pt(1.7 nm)/Nb(30 nm)/Ni$_8$Fe$_2$(6 nm)/Nb(30 nm)/Pt(1.7 nm)/Co(0.8 nm)/Pt(2.0 nm) device. **c**, Normalized intensity of the real part of $\Delta S_{12}$ across $T_c$ for the Pt/Co(0.8 nm)/Pt-absent device with $d = 10−25$ µm. **d**, Data equivalent to **c** but for the Pt/Co(0.8 nm)/Pt-present device. Each inset shows the associated $d$-dependence of the signal intensity above and below $T_c$. The dashed lines in **c,d** are guide to the eyes whereas the solid lines in each inset are fits to an exponential decay function to estimate the SW attenuation length $\lambda_{att}$[33,34] (see Methods).

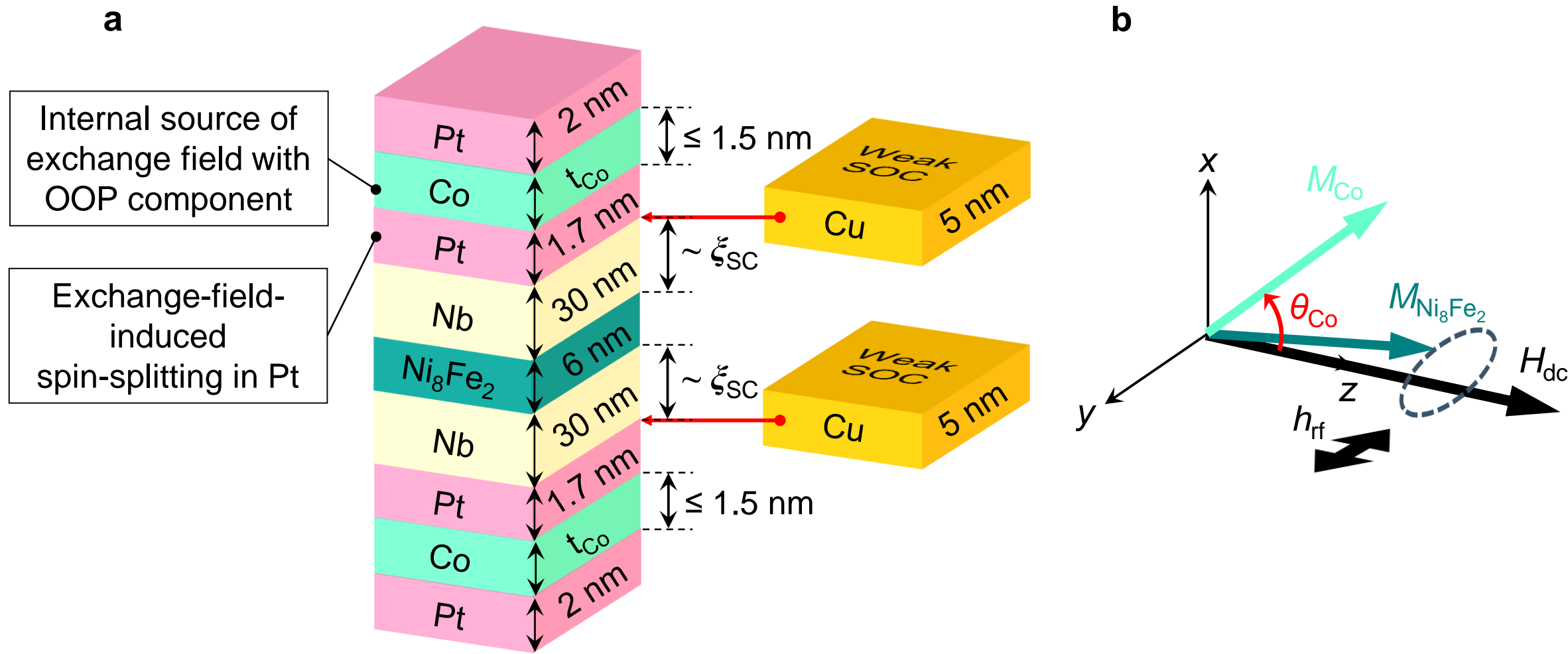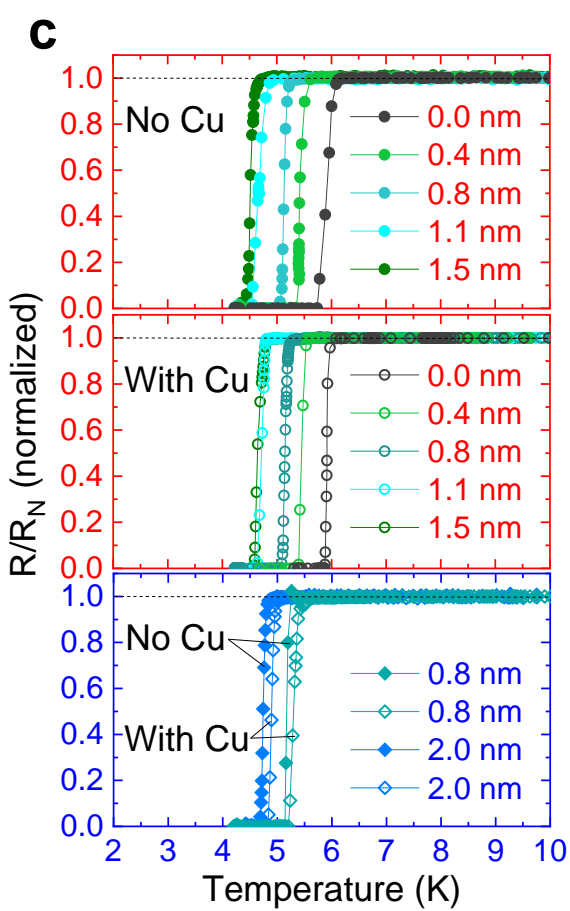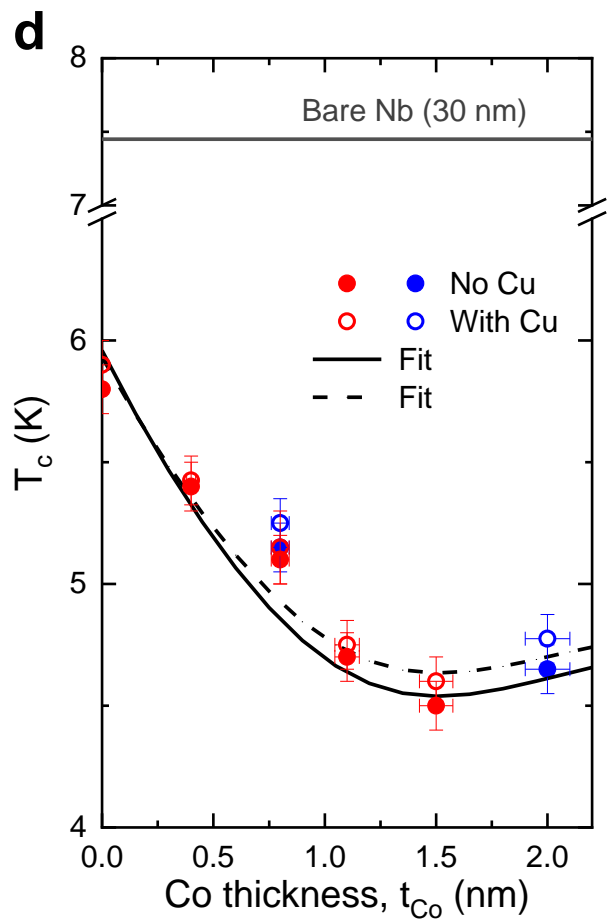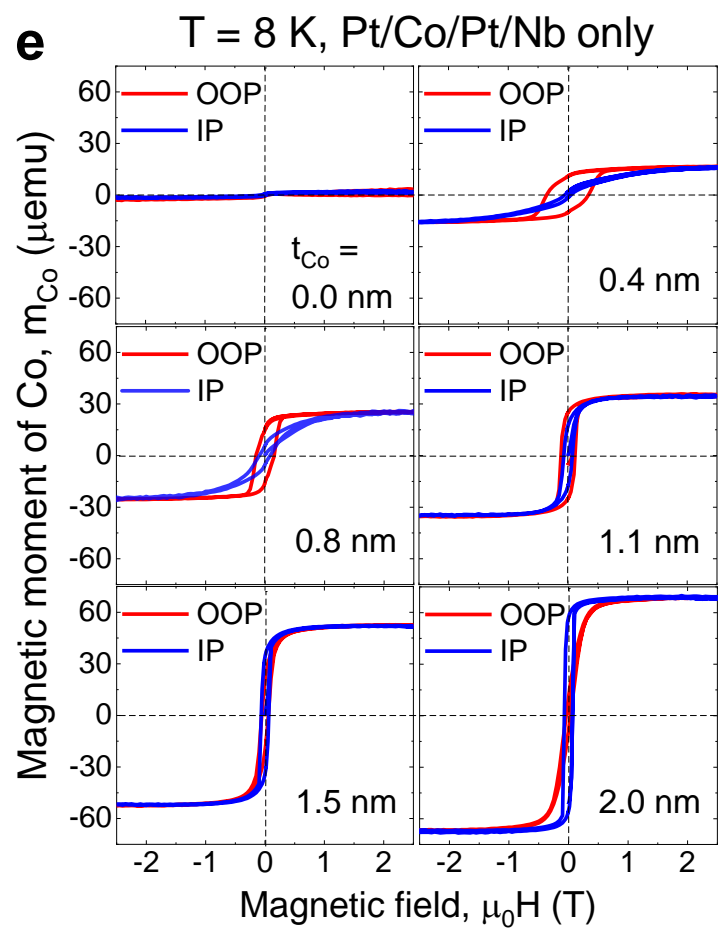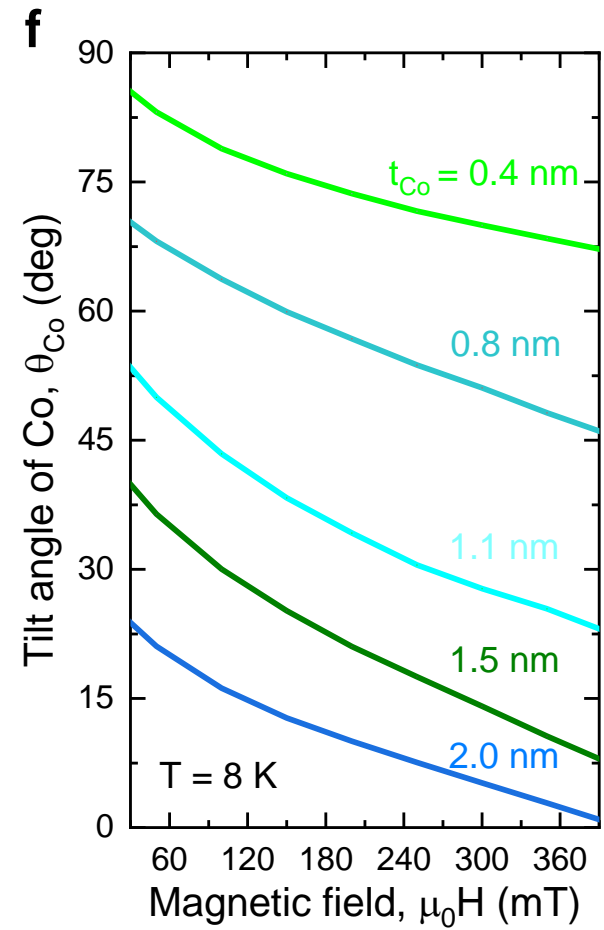

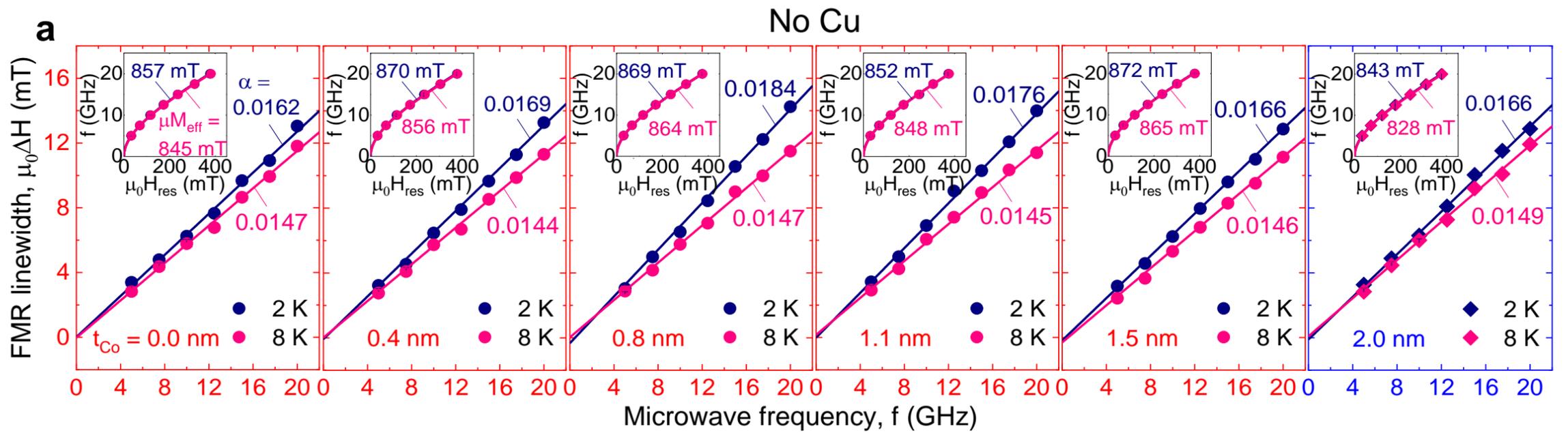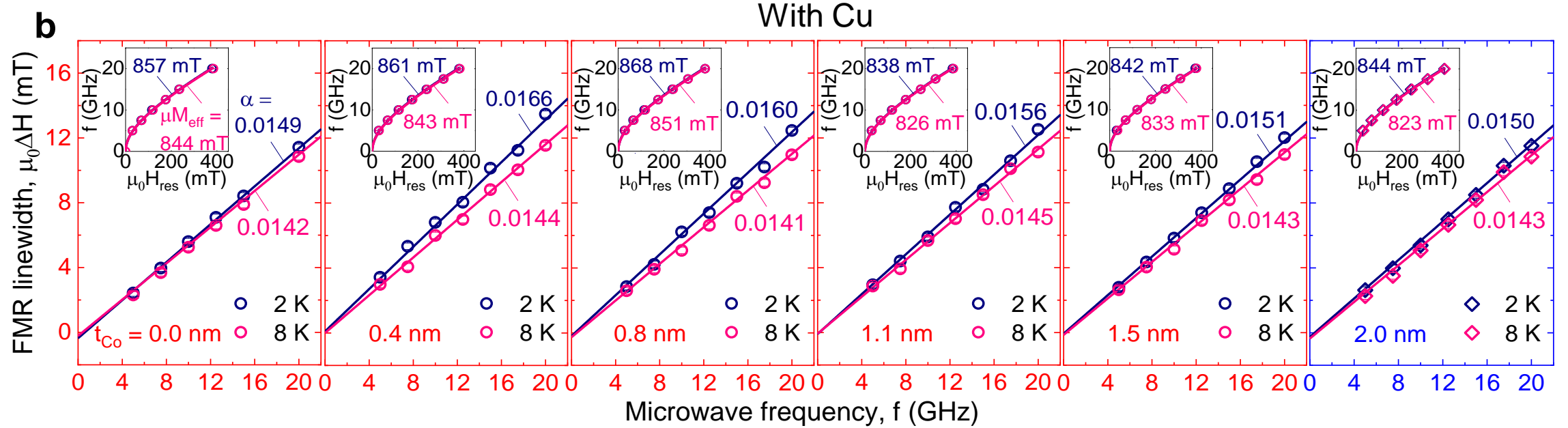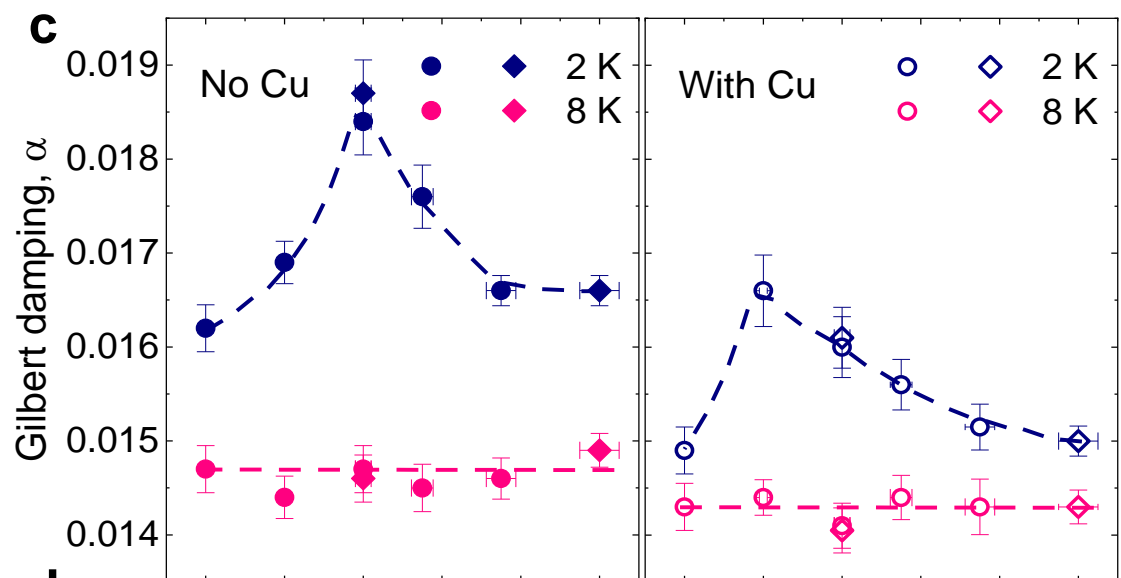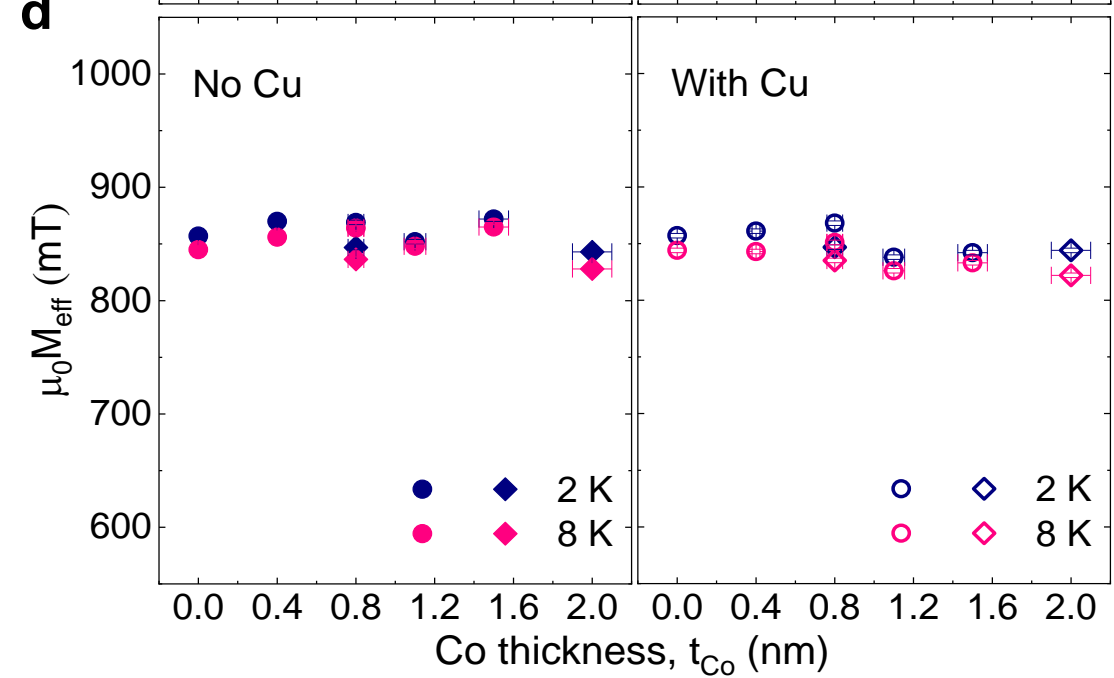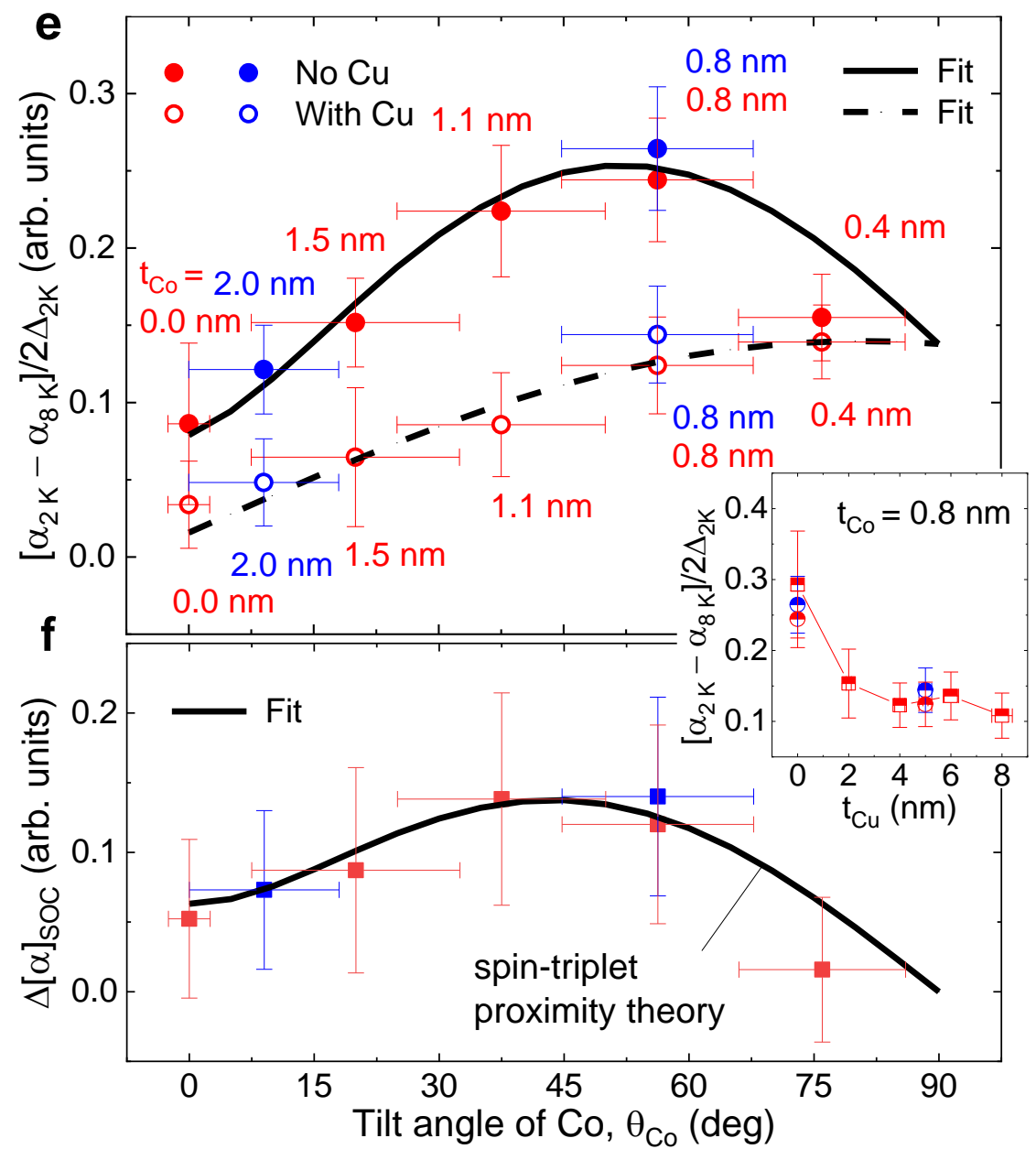

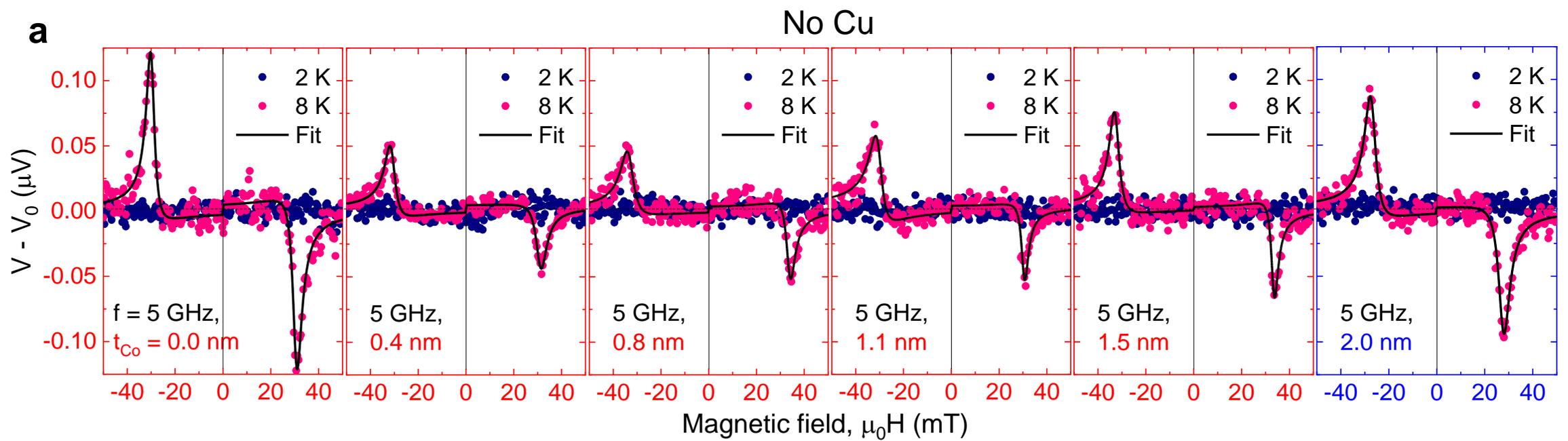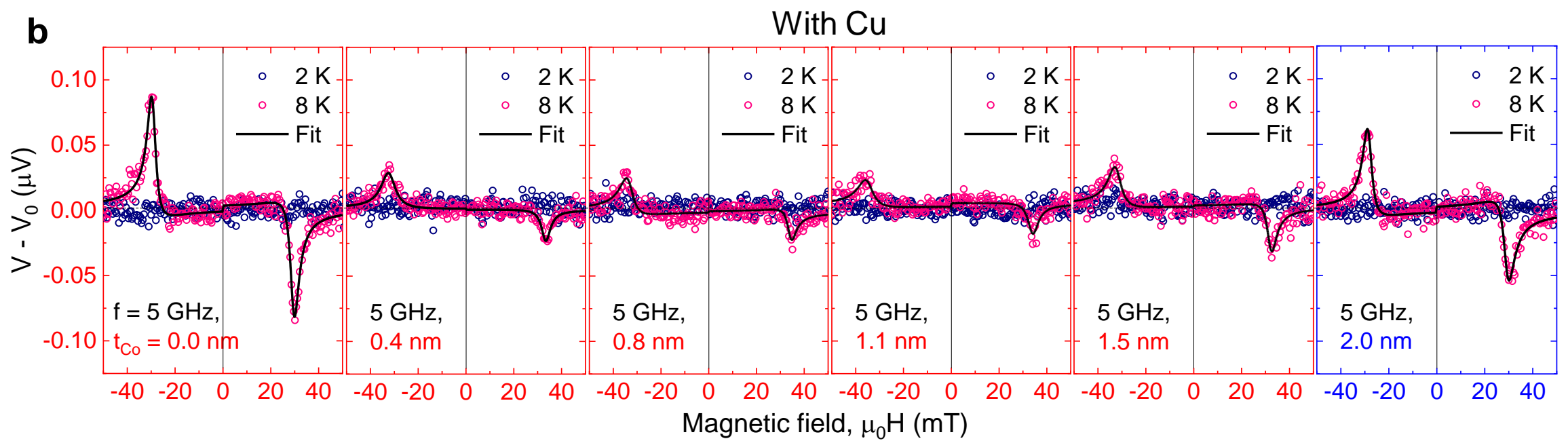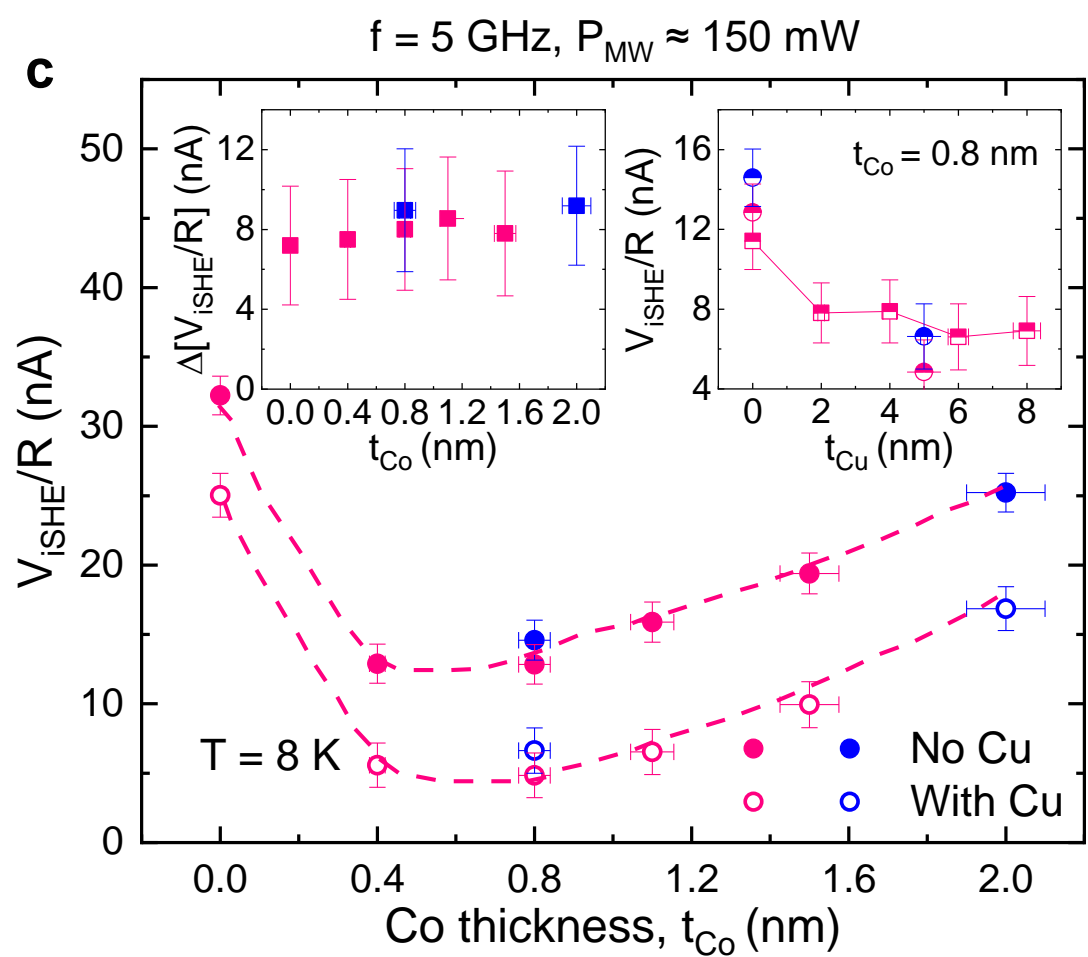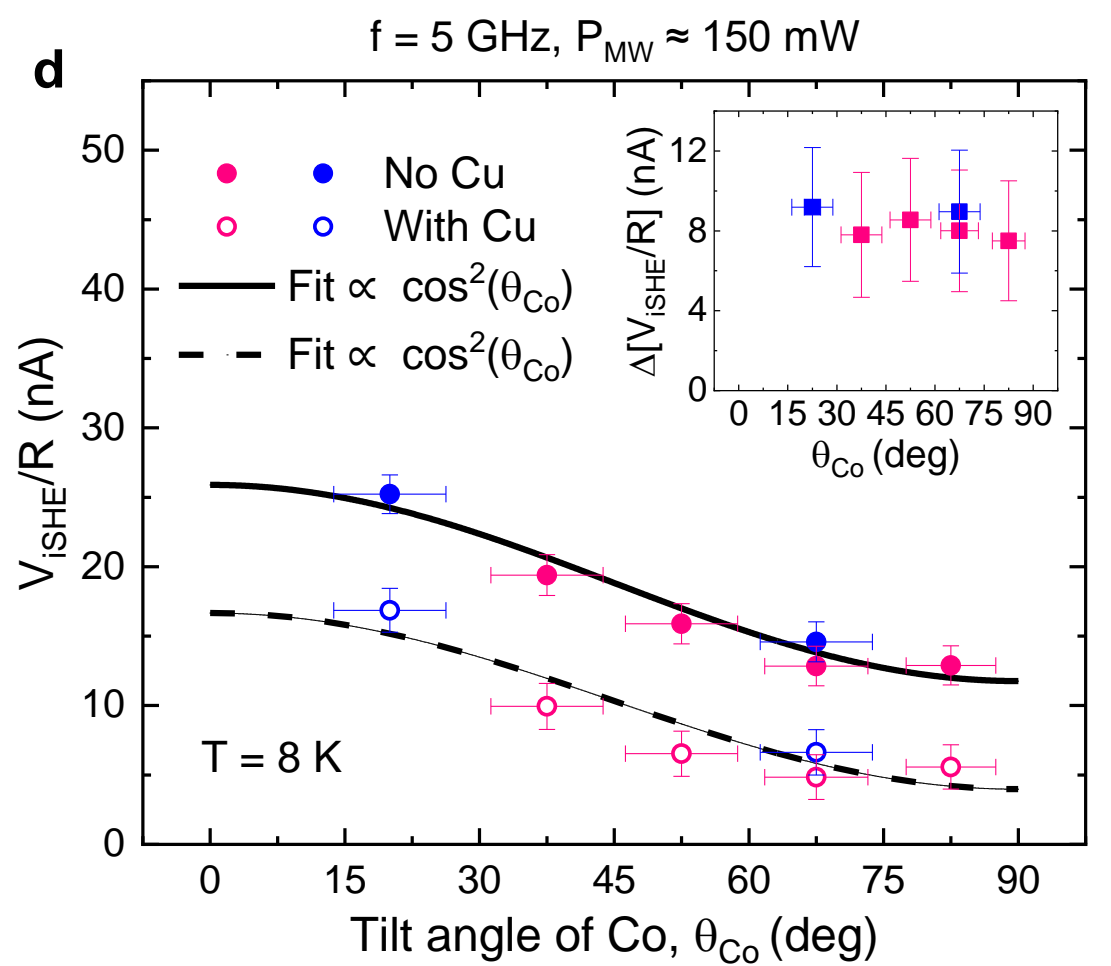

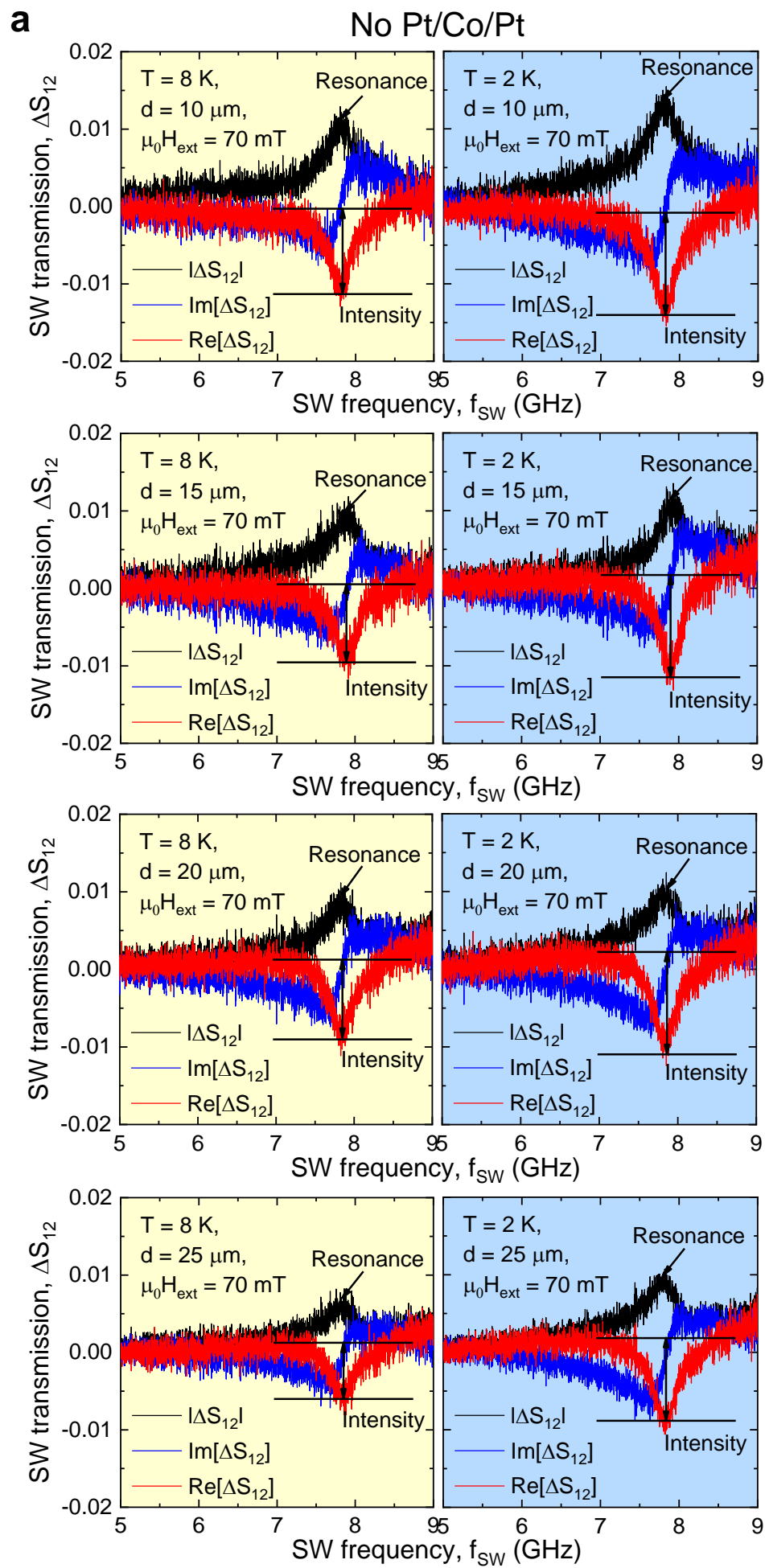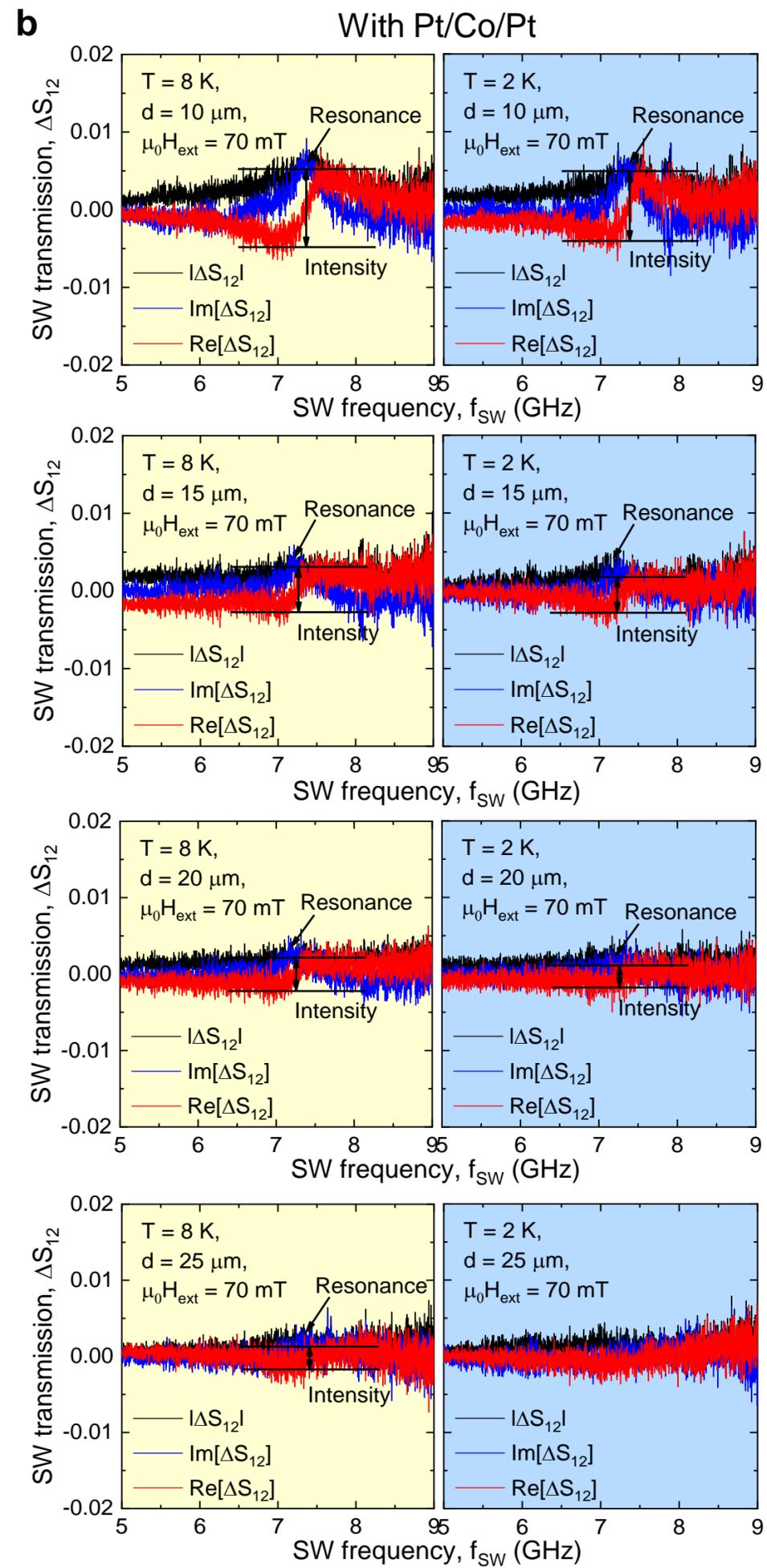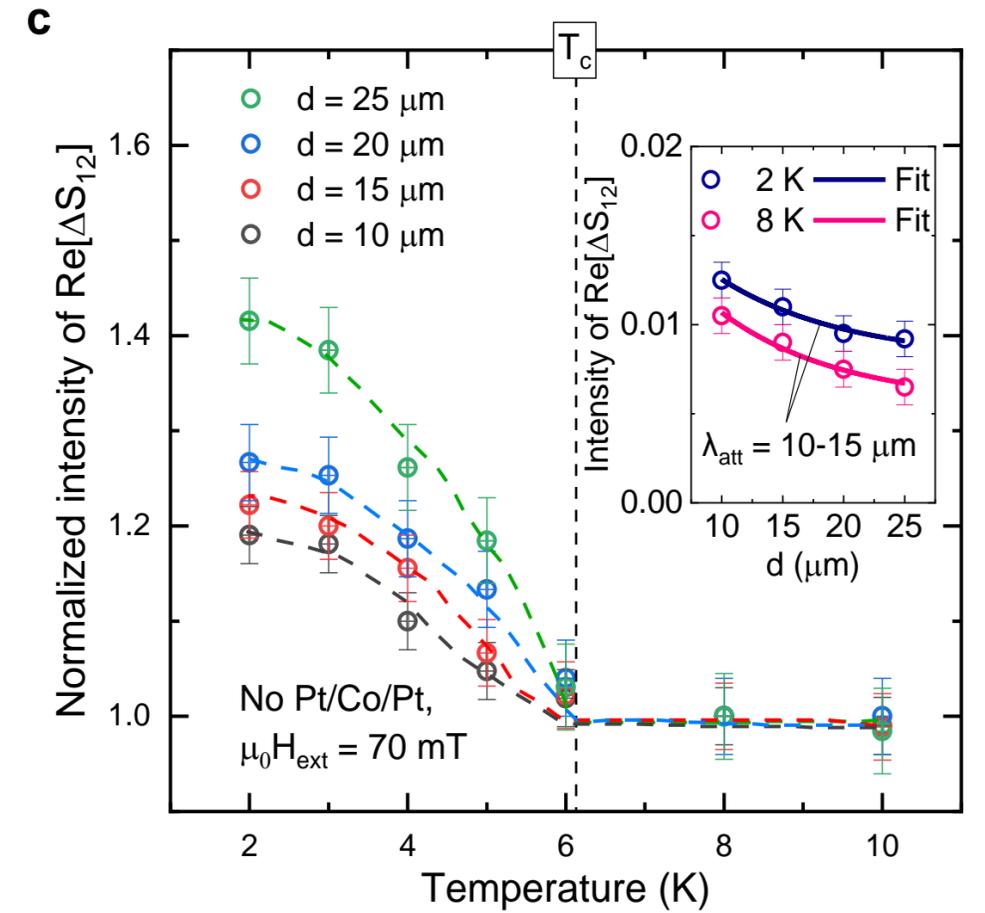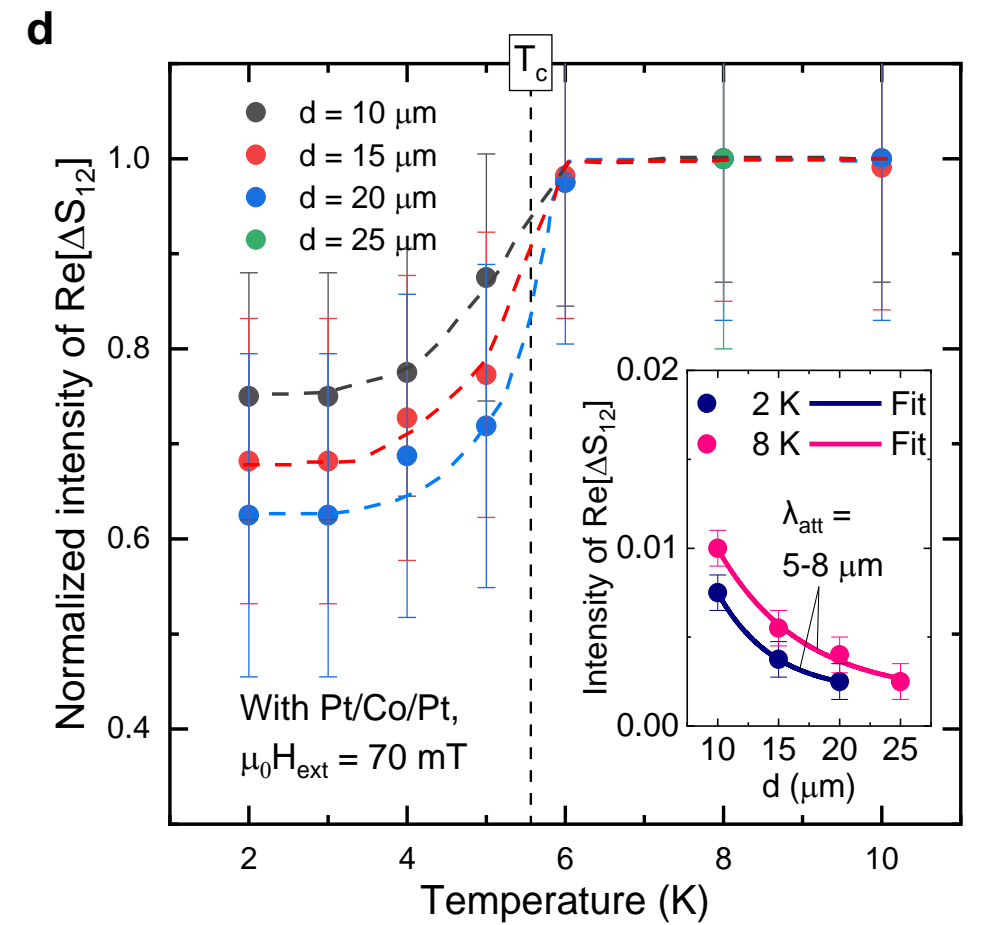